\documentstyle[12pt]{article}
\include{epsf}
\begin{document}
\title{\Large \bf    Evolution of economic entities under
heterogeneous political/environmental conditions within a
Bak-Sneppen-like dynamics}

\author{ Marcel Ausloos$^1$, Paulette Clippe$^2$ and Andrzej
P\c{e}kalski$^3$ \\ $^1$ SUPRAS(\footnote{a member of SUPRATECS})
and GRASP, B5, Sart Tilman Campus, \\B-4000 Li$\grave e$ge,
Euroland, \\$^2$ GRASP, B5, Univ. de Li$\grave e$ge, \\B-4000
Li$\grave e$ge, Euroland, \\$^3$ Institute of Theoretical Physics,
University of Wroc{\l}aw,
\\pl. Maxa Borna 9, \\PL-50-204 Wroc{\l}aw, Poland }

\maketitle

 \vspace{6ex}
\centerline{ email:} Marcel.Ausloos@ulg.ac.be,
Paulette.Clippe@ulg.ac.be, apekal@ift.uni.wroc.pl

\newpage

\begin{abstract} A model for  economic behavior, under
heterogeneous spatial economic conditions is developed. The  role
of selection pressure in a Bak-Sneppen-like dynamics with entity
diffusion on a lattice is studied by Monte-Carlo simulation taking
into account business rule(s), like enterprise - enterprise short
range location "interaction"(s), business plan(s) through
spin-offs or merging and enterprise survival evolution law(s). It
is numerically found that the model leads to a sort of phase
transition for the fitness gap as a function of   the selection
pressure.

\end{abstract} \vskip 0.5cm

\noindent Pacs: 89.65.Gh, 05.10.Ln,  89.75.-k, 07.05.Tp,
05.65.+b


\section{Introduction}

In a previous paper \cite{ACPbali} we have raised the
question whether one could through modern econophysics ideas
and theories \cite{baliproceed} touch upon questions related
to changing economic world  conditions, e.g. globalization
or mondialisation and  delocalization, and  discuss the
consequences of such politics. The Berlin wall destruction
and opening of markets in Eastern Europe and Central Asia to
so called liberal economy was said to be similar to an
increase in "physical volume or available space". The
disparity in "economic conditions", like the tax systems, or
different  workers' skills or wages, weather, available
information, ... seemed to be also similar to ''external
field''  conditions in condensed matter.

In a first approach we have asked whether the evolution of
the concentration  of enterprises, their spatial
distribution, their (so we say) "fitness" under varying in
time and space economic field conditions can be derived from
 some microscopic-like model as often found in statistical
physics to describe complex phenomena. We have raised the
question whether macroeconomy seems to work as a
self-organized system \cite{Bak}, characterized by scaling
laws, whether models and theories pertaining to such
features characterized by punctuated equilibrium contain
cycles or stable states \cite{cycleK,cycleKS0,januszma1}.

In \cite{ACPbali}  we have introduced an economic  world
model  as a lattice divided into $\nu$ (= 3) regions  of
equal size. At the origin of time all "firms" (enterprises,
agents, ..)  are located only in region  I. A barrier
preventing companies to enter  regions II and  III is
removed at a given $t_1$ time. Each company ($i$) is
characterized by one real number  $f_i$, belonging to the
[0,1] interval. Each region is under some economic
conditions, an "external field", represented by $one$  real
number,  $F$ also belonging to [0,1]. The best condition,
"by symmetry", is  $F = 0.5$. We  have searched for the
evolution of concentrations in regions II and III as invaded
according to a simple diffusion rule. A so called business
plan, based on a biased merging and spin-off creation
alternative was  considered for the enterprise evolution.
The $f$ value of the new firms was considered to be obtained
according to various types of memories depending on the $f$
of the company parents \cite{ACPbali,ACPevol}.

In such a scheme, some attempt was made to connect
macroeconomy and econophysics \cite{reconcile} through
questions concerning whether enterprises survive or not, get
better or worse, how the diffusion front  moves toward
regions II and III, etc.  One ingredient of the Monte-Carlo
algorithm  was to consider that the evolution  was randomly
driven, in the sense that the entity picked up by the
Monte-Carlo procedure was independent of the position and
$f$ value of the firm.

However some other dynamics can be imagined. In particular recall
the Bak-Sneppen evolution model \cite{BakSneppen} which has
considered that species being assigned a so called fitness $f$
have a better chance of surviving if $f$ is large. The population
evolution is controlled through the entity with the minimal
fitness at every discrete time step. This species is replaced by
a new one, with a new fitness, taken from a uniform distribution
on [0,1]. Moreover an interaction is introduced, e.g. by
(randomly) modifying the fitness of the neighbors of the chosen
entity. This interaction represents a co-evolution of related
species. Many variants  have been studied, including an
anisotropic Bak-Sneppen model \cite{HeadRodgers}, a tree growth
\cite{VdWMASOC} evolution with or without species screening
\cite{VDWMAK,VvWMAscreening}, with or without long range
interactions \cite{VDWMAHV}.... Applications of the Bak-Sneppen
ideas occur in many fields, from biological problems, like
bacteria colony evolution \cite{BSKovalev,BSDonangelo}  to
macro-economical processes \cite{BSecon1Cuni,BSecon2Yamano}.

We reformulate the simple model presented in \cite{ACPbali} for
the evolution of economic entities under varying economic
conditions introducing a Bak-Sneppen- like dynamics. In Sect. 2,
we present  the new model algorithm which therefore stresses the
role of the selection pressure and the Bak-Sneppen-like dynamics.
In section 3, we outline a few results, like the diffusion front
penetration, and the concentration of entity evolution. It is
numerically observed that the model leads to a phase
transition-like scenario for the fitness gap as a function of the
selection pressure.  A short conclusion is to be found in Sect. 4.

\section{Model and Monte Carlo Simulation Algorithm}

We consider a square symmetry lattice of linear dimensions
$L_x \times L_y$. The $x$ segment is divided into 3 parts
(regions I, II and III) of equal size ($L_x$ =150,  $L_y =
201$). The entities we consider, called thereafter {\it
firms}, are located on the lattice sites. A site may be
either empty or occupied by one firm.

Each firm is characterized by its location on the lattice and a
number $f \in [0,1]$.  All external conditions influencing the
dynamics of the firms (labor conditions, fiscal system,
availability of human and natural resources, etc) are summarized
in a single value $F \in [0,1]$ - the external field. Since the
agreement between $f$ and $F$ will determine the firm's survival,
see below, we shall call $f$ the {\it fitness} of the firms.

Initially all firms are located at random positions in
region I  with initial density $c(0)$ and with random values
of the $f$'s. Like in the Bak and Sneppen extremal dynamics
model \cite{BakSneppen} we choose a firm with the lowest
appealing fitness from the field point of view, do not
remove the firm immediately from the system as was done in
the Bak and Sneppen model, but check the firm's survival
chance. This  depends on the difference between the fitness
of the firm and the external field and the selection
pressure. If the firm does not survive the check, it is
removed  from the system and all firms which are the nearest
neighbors (von Neumann neighborhood) of the chosen firm
receive new, random values for their fitness.

If not removed, the firm may change position, merges with
another firm  or creates spin-off(s).

After a certain time the border between the region I and
region  II is opened, thus permitting the motion of firms
into regions II and III. At the same time the external field
changes in region I, assuming a new value $F_I$, different
from that in region  II ($F_{II}$) and III ($F_{III}$).
There is only one such change of the field  (in contrast to
cases examined in \cite{ACPbali}).

To complete a Monte Carlo Step (MCS) one has to pick as many
firms as there were at the beginning of that step.

The algorithm goes as follows:
\begin {enumerate}
\item The firm ($i$) which has its fitness farthest away from the
field value is picked. The search is made in the whole system, and
the fitness is compared to the region field value, i.e. the
fitness of a firm in the region I is compared with $F_I$, that in
the region II with $F_{II}$, etc.

\item  The survival probability is calculated as

\begin{equation} p_i \,= \,\exp(-sel|f_i - F|) , \end{equation}

and checked against a random number $r_i \in [0,1]$ taken from a
uniform distribution. If $r_i > p_i$, the firm is removed from the
system and its nearest neighbors are given new, arbitrary, but
within the [0,1] interval, values of  fitness,

\item If the firm survived, then 7 random searches are made
in the von Neumann neighborhood (4 sites) for an empty
place. As soon as the search is successful, the firm moves
there.

\item A random search is next made  for a partner in the von
Neumann neighborhood of the new position. If found at the site
$j$ then

\item with a probability 0.01 the two firms merge, creating
a new firm at the location of the first one, with a new
fitness given by

\begin{equation} f_i = \frac{1}{2}\left[(f_i + f_j) +
sign[0.5 - r] |f_i - f_j|\right],
 \end{equation}
 where  $r$  is a
random number    in [0,1]. The second firm is eliminated. If
the new fitness for the (first) firm is greater than 1, the
firm is eliminated too.

\item with a probability 0.99 the two firms produce a new ($k$)
firm (spin-off) with a fitness  given by the formula
\begin{equation} f_k = \frac{1}{2}\left[(f_i + f_j) +
sign[0.5 - r] |f_i - f_j|\right],
 \end{equation}
 15 random searches in the Moore neighborhood (9 sites) of the
first firm are made to put the spin-off firm there at the first
found empty site. If the attempts are unsuccessful, the new firm
is not created.

\end{enumerate}

After choosing all agents, one MC time step is done. We have
in such a way limited our investigations to 1500  MCS.

In each case, we have started with an initial concentration $c$ =
0.8 in region I, and have destroyed the  wall  between region I
and II after 100 MCS. The external field takes values 0.3, 0.5,
0.6 in the three regions after $t_1$ and remains constant
thereafter. The results reported in Sect. 3 refer  to averages
over ten runs.

\section{Results}

In the following we stress results pertaining to cases
demonstrating the newness of the model with respect to our
previous work \cite{ACPbali}.

In Fig. 1,  the number of firms existing in the three regions is
shown as a function of time for a $sel$ = 0.5. Recall that there
are  10050 lattice sites in each region. It is observed that there
is a rapid  decrease in concentration in the first region as soon
as the I-II wall is open, but the death and birth process
stabilizes the concentration at nearly $c$ = 0.4   after a few
Monte-Carlo steps, even though the concentrations in regions II
and III still increase due to the largely available space.

It is unclear whether for very long time the concentrations
would level off to the value of region I or to other values
in regions II and III.  However an interesting qualitative
information is seen through the position of the diffusion
front (going, recall from left to right, in this world) for
various $sel$ values. (Fig.2).  The break toward some
stationary-like asymptotic concentration value in region I,
as seen in Fig. 1, seems to occur at the same time as the
front progresses, as seen in Fig. 2; the break seems to mark
  an apparently different growth law. We
have not studied the diffusion coefficient, but the
difference in behavior might be traced to the effects of
diffusion (together with the birth and death) process  in
presence of (moreover here mobile) barriers, as in
\cite{PAdiff147,PAdiff195}. The front propagation for $sel
<< 0.65$ seems to be a simple Brownian like process.

The complete spatial distribution and its evolution in time would
require a movie like display; instead we give snapshots for the
value of the concentration along vertical columns,  for a few
Monte-Carlo times  and two $sel$ values in Fig. 3(a-b)
respectively. Recall that sharp field gradients occur at $x = 50$
and $100$. For the low $sel$ (= 0.3) the  concentration in region
I hardly varies with time (and column position), and remains  near
0.4, as can be hinted from Fig.1 for $sel$ = 0.5.  After
sufficient Monte-Carlo steps the concentration seems to  level off
in region II, although it is not clear what would be the final
concentration at very large MC time.  For ''large'' $sel$ value
the concentration in region I is markedly decreased, and is quasi
nonexistent in regions II and III.

It is therefore of interest to emphasize the role of $sel$
at an intermediary time, as in Fig. 4, where it is seen that
 the behavior does not much differ from low to high $sel$
except for the concentration  amplitude, - which seems to
decrease linearly with $sel$ in region I.

The fitness  value evolution  in the three regions  is not
spectacular since it ''rather quickly'' reaches the optimal one,
as constrained by the external field. The fitness optimal value in
each region is easily reached and remain stable with MCS. It is
more interesting to display the behavior of the fitness in line
with the Bak-Sneppen dynamics. Recall that the interest of the
dynamics is, beside the entity interactions, the choice of the
less fit for starting an avalanche process. Therefore we have
calculated the number (N) of times the less fit entity is reached
(picked), and its fitness value, during the process. The data for
N as a function of $f$ in the case $sel$ = 0.5 is shown in Fig. 5
on a semi-log plot.

 It is observed that the
number of unfit entities is large and about the same in the
three regions. Moreover a rather well defined gap occurs in
the distribution of ''picked firms'', - a gap centered on
the external field value. This $f_{gap}$ can be displayed in
Fig. 6 as a function of $sel$. Except for the asymmetry due
to the external field value, the behavior of the gap  can be
reconciled with what is expected for the density at
liquid-gas transitions as a function of temperature. The
finite size of the system is likely the cause for inducing a
non zero gap at large $sel$. A critical $sel$ value near
0.66 is found whatever the external field.

\section{Conclusions}

In summary, we have adapted a birth-death-diffusion process
of macroeconomic evolution  with a Bak-Sneppen-like
dynamics. From this set of results,  we have observed that
there are relatively well marked effects due to the
"selection pressure",  including a ''critical value''
reminiscent of the critical temperature at second order
phase transitions. The constraining economic ("external")
field implies stable concentration distributions, as far as
examined. The diffusion process is not spectacular,
apparently being Brownian like. We have not searched for
avalanches as found in self-organized processes.

Result robustness should be further checked with respect to
the parameters which are involved : number of regions,
lattice size(s), lattice symmetry, initial concentration(s),
 field time sequence(s), and   time for $F$ field changes,
selection pressure, hopping distance for the diffusion
process, number of 'spin off's, the latter creation
probability, ...

Further improvements  can be also suggested both from a
macro and micro economy point of view, as well as from
physical system studies. A company cannot be described by
one scalar number $f_i$, but a vector model  coupled to a
vector field should be more realistic. Moreover the role of
the business plan through the birth and death process
description mapping merging and spin off processes is also
to be improved.  The fitness distribution and evolution
law(s) might also be changed for better reflecting
macroeconomy findings. Analytical work could be of interest
to search for bifurcation and chaotic  conditions, if any.

\vskip 0.6cm

{\bf Acknowledgments}

\vskip 0.6cm

MA and AP thank the CGRI and KBN for partial financial
support allowing mutual visits during this work process. MA
and PC also thank  an Action Concert\'ee Program of the
University of Li$\grave e$ge (ARC 02/07-293). 

\newpage

{ \Large{\bf Figure captions}} \vskip 0.5cm

{\bf Figure 1} -- Average (over 10 runs) number of firms existing
in the three regions as a function of time for $sel$ = 0.5 and
field values $F$ =0.3, 0.5, 0.6 after $t_1$=100 MCS.

\vskip 0.5cm {\bf Figure 2} -- Position of the diffusion front for
various $sel$ values with external field values $F$ = 0.3, 0.5,
0.6 after $t_1$ = 100 MCS in the three regions respectively.

\vskip 0.5cm {\bf Figure 3} -- The  concentration  in the three
regions for a few MCS   when the  external field takes values $F$
=0.3, 0.5, 0.6 after $t_1$=100 MCS for $sel$ = (a) 0.3 and (b)
0.8.

\vskip 0.5cm {\bf Figure 4} -- The  concentration in the three
regions at t = 1000 MCS, for external field values $F$ = 0.3, 0.5,
0.6 for   different  $sel$ values.

\vskip 0.5cm {\bf Figure 5} -- Semi-log plot for the number N of
(picked) ''less fit entities'' (as defined in the text) in the
three regions after 1500 MCS when the external field values $F$ =
0.3, 0.5, 0.6 respectively  for $sel$ = 0.5.

\vskip 0.5cm {\bf Figure 6} --  Display of  $f_{gap}$, i.e. gap
in picked $f$ values, as a function of $sel$ in the three regions
when the external field values are $F$ = 0.3, 0.5, 0.6
respectively.

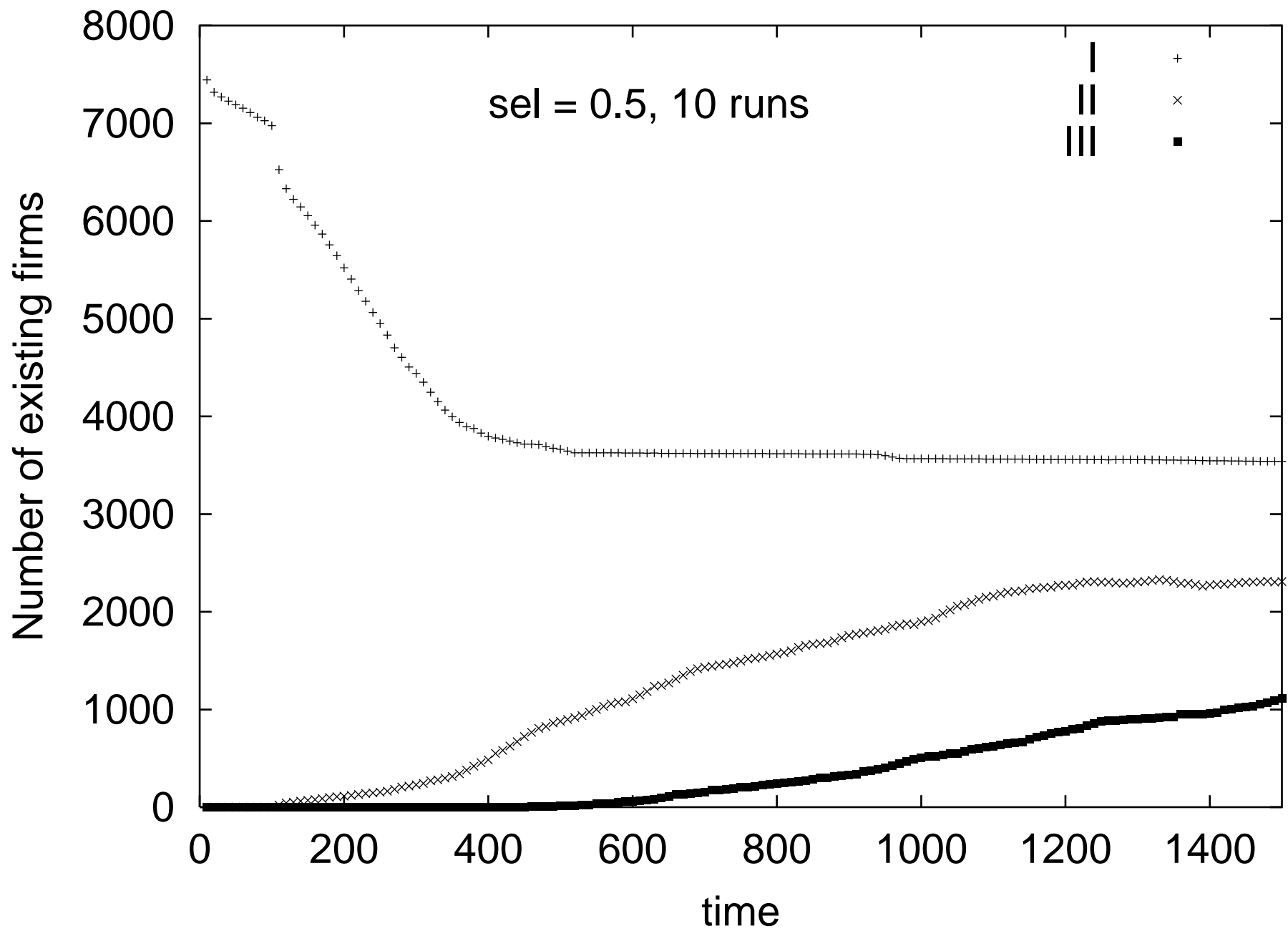
\end{document}